%% file: paper_arxiv.tex
\documentclass[runningheads]{llncs}                     % onecolumn (standard format)
\usepackage{etex}
\usepackage{picins}
\usepackage{graphicx}

\usepackage{amssymb,amsmath}

\usepackage[show]{ed}
\usepackage{amstext}
\usepackage{xspace}
\usepackage{tikz}
\usetikzlibrary{arrows}
\usetikzlibrary{shapes}
\usetikzlibrary{positioning}
\usetikzlibrary{calc}
\usepackage{amssymb}
\usepackage{url}
\usepackage{paralist}
\usepackage{xy}
\xyoption{v2}
\xyoption{curve}
\xyoption{line}
\usepackage{tabularx}
\usepackage{picins}

\newcommand{\DG}{\ensuremath{{\mathcal{D}}}}

\newcommand{\DGRoots}[1]{\ensuremath{\lceil #1\rceil}}

\newcommand{\cS}{\ensuremath{{\cal S}}}

\newcommand{\calL}{{\mathcal{L}}}

\newcommand{\glinka}[3]{\!\xymatrix{{#1} \ar@{=>}[r]^{#2} & {#3}}\!\!}
\newcommand{\longglinka}[3]{\!\xymatrix{#1 \ar@{=>}[rr]^{#2} && #3}\!\!}

\newcommand{\hlinka}[3]{\!\xymatrix{#1 \ar@{=>}[r]^{#2}_{\mathit{hide}} & #3}\!\!}
\newcommand{\longhlinka}[3]{\!\xymatrix{#1 \ar@{=>}[rr]^{#2}_{\mathit{hide}} && #3}\!\!}
\newcommand{\flinka}[3]{\!\xymatrix{#1 \ar@{=>}[r]^{#2}_{\mathit{free}} & #3}\!\!}
\newcommand{\longflinka}[3]{\!\xymatrix{#1 \ar@{=>}[rr]^{#2}_{\mathit{free}} && #3}\!\!}

\newcommand{\tglinka}[3]{\!\xymatrix{#1 \ar@{==>}[r]^{#2} & #3}\!\!}
\newcommand{\longtglinka}[3]{\!\xymatrix{#1 \ar@{==>}[rr]^{#2} && #3}\!\!}
\newcommand{\thlinka}[4]{\!\xymatrix{#1 \ar@{==>}[r]^{#2}_{{\mathit{hide}}\ #3} & #4}\!\!}
\newcommand{\longthlinka}[4]{\!\xymatrix{#1 \ar@{==>}[rr]^{#2}_{{\mathit{hide}}\ #3} && #4}\!\!}
\newcommand{\tflinka}[4]{\!\xymatrix{#1 \ar@{==>}[r]^{#2}_{\mathit{free}\ #3} & #4}\!\!}
\newcommand{\tllinka}[3]{\!\xymatrix{#1 \ar@{-->}[r]^{#2} & #3}\!\!}

\newcommand{\gclinka}[3]{\!\xymatrix{#1 \ar@{=>}[r]^{#2}_{\mathit{cons}} & #3}\!\!}
\newcommand{\gdlinka}[3]{\!\xymatrix{#1 \ar@{=>}[r]^{#2}_{\mathit{def}} & #3}\!\!}
\newcommand{\gmlinka}[3]{\!\xymatrix{#1 \ar@{=>}[r]^{#2}_{\mathit{mono}} & #3}\!\!}

\newcommand{\tclinka}[3]{\!\xymatrix{#1 \ar@{==>}[r]^{#2}_{\mathit{cons}} & #3}\!\!}
\newcommand{\tdlinka}[3]{\!\xymatrix{#1 \ar@{==>}[r]^{#2}_{\mathit{def}} & #3}\!\!}
\newcommand{\tmlinka}[3]{\!\xymatrix{#1 \ar@{==>}[r]^{#2}_{\mathit{mono}} & #3}\!\!}

\newcommand{\longtclinka}[3]{\!\xymatrix{#1 \ar@{==>}[rr]^{#2}_{\mathit{cons}} && #3}\!\!}
\newcommand{\longtdlinka}[3]{\!\xymatrix{#1 \ar@{==>}[rr]^{#2}_{\mathit{def}} && #3}\!\!}
\newcommand{\longtmlinka}[3]{\!\xymatrix{#1 \ar@{==>}[rr]^{#2}_{\mathit{mono}} && #3}\!\!}

\newcommand{\greaches}[1]{\!\xymatrix{\ar@{)=>}[r]^{#1} & }\!\!}
\newcommand{\lreaches}[1]{\!\xymatrix{\ar@{)->}[r]^{#1} & }\!\!}

\newcommand{\loc}{\mathit {loc}}
\newcommand{\Dom}[1]{\mathit {Dom}_{#1}}
\newcommand{\dom}{\mathit {dom}}
\newcommand{\Imports}[1]{\mathit {Imports}_{#1}}

\newcommand{\derives}{\ensuremath{\vdash}} 
\newcommand{\stroke}{|}

\newcommand{\Mod}{\mathbf{Mod}}
\newcommand{\Sen}{\mathbf{Sen}}

\newcommand{\Sign}{\mathbf {Sign}}

\newcommand{\presigma}{{\sigma}}
\newcommand{\pretau}{{\tau}}
\newcommand{\pretheta}{{\theta}}

\newcommand{\symbols}{\ensuremath{\mathit{sym}}}
\newcommand{\supports}{\ensuremath{\sqsubset}}

\newcommand{\SIGExtension}[2]{\langle #1\rangle_{#2}}

\newcommand{\signature}{\textit{sig}}
\newcommand{\axioms}{\textit{ax}}
\newcommand{\lemmas}{\textit{lem}}

\newcommand{\Signature}[1]{\textit{Sig}_{\cal #1}}
\newcommand{\Axioms}[1]{\textit{Ax}_{\cal #1}}
\newcommand{\Lemmas}[1]{\textit{Lem}_{\cal #1}}

\newcommand{\Support}{\textit{Supp}}

\newcommand{\AllAxioms}{\Axioms{}}
\newcommand{\AllLemmata}{\textit{Lem}}

\tikzstyle{globallink}=[double,->]
\tikzstyle{globalreachable}=[double,latex' reversed-latex',transform shape]
\tikzstyle{globaltheoremlink}=[double,dashed,->]
\tikzstyle{locallink}=[thin,->]
\tikzstyle{localreachable}=[thin,latex' reversed-latex']
\tikzstyle{localtheoremlink}=[thin,->,transform shape]
\tikzstyle{hidinglink}=[->]
\tikzstyle{hidingtheoremlink}=[->]

\tikzstyle{namednode}=[ellipse,draw,fill=black!10]

\input{defs}

\begin{document}

\title{Structure Formation in Large Theories%\thanks{Grants or other notes
\thanks{The final publication is available at http://link.springer.com as
part of the proceedings of the Conference on Intelligent Computer Mathematics 2015.}
}

% \subtitle{Do you have a subtitle?\\ If so, write it here}


\author{Serge Autexier \and Dieter Hutter}


\institute{German Research Center for Artificial Intelligence \\
           Bibliothekstr. 1, 28359 Bremen, Germany \\
           \email{\{serge.autexier$\mid$dieter.hutter\}@dfki.de}}
           
%\institute{F. Author \at
%              first address \\
%              Tel.: +123-45-678910\\
%              Fax: +123-45-678910\\
%              \email{fauthor@example.com}           %  \\
%%             \emph{Present address:} of F. Author  %  if needed

\date{Received: date / Accepted: date}

\maketitle

\allowdisplaybreaks

\begin{abstract}
  Structuring theories is one of the main approaches to
  reduce the combinatorial explosion associated with reasoning and
  exploring large theories. In the past we developed the notion
  of development graphs as a means to represent and
  maintain structured theories. In this paper we present a
  methodology and a resulting implementation to reveal the 
  hidden structure of flat theories by transforming
  them into detailed development graphs. We review
  our approach using plain TSTP-representations of 
  MIZAR articles obtaining more structured and also more
  concise theories.
\end{abstract}

\section{Introduction}
\label{sect:introduction}
It has been long recognized that the modularity of specifications is an indispensable prerequisite 
for an efficient reasoning in complex domains. Algebraic specification techniques provide 
appropriate frameworks for structuring complex specifications and the authors introduced 
the notion of an development graph \cite{Hutter00a,AH05,MAH-05-a} as a technical means to work 
with and reason about such structured specifications. While its use presupposes the development
of theories having the intended structures already in mind, there are various applications of Formal Methods 
in which theories are automatically generated in an entirely unstructured representation. Thus, there
is a need for a computer-aided structure formation for large theories, which allows for an efficient
reasoning in such theories.

In this paper we present an initial approach to support structure formations in large 
unstructured specifications. The idea is to provide a calculus and a corresponding methodology 
to crystalize intrinsic structures hidden in a specification and represent them explicitly in terms
of development graphs. Step by step, the specification is split into different nodes
resulting in increasingly richer development graphs. On the opposite, common concepts
that are scattered in different specifications are identified and unified in a common theory.

We start with a discussion on syntactical properties to measure the appropriateness of a structuring 
and specify invariants underlying a structure formation process. Based on this general framework
we present a calculus (and heuristics to guide this calculus) to transform development graphs in order
to enrich the explicitly given structure. We review our framework with the help of the Mizar Mathematical 
Library (\url{http://www.mizar.org/}) providing hundreds of articles which are subject to our structure formation process.

\section{Development Graphs for Structure Formation}

We base our framework on the notions of development graphs (and thus on the notion
of institutions \cite{GoguenBurstall92}) to specify and reason about
structured specifications. Development graphs ${\DG}$ are acyclic, directed graphs 
$\langle {\cal N}, {\cal L} \rangle$, the nodes ${\cal N}$ denote individual
theories and the links ${\cal L}$ indicate theory inclusions with respect to signature 
morphisms attached to the links.  Each node $N \in \calN$ of the graph 
is a tuple $(\signature^N, \axioms^N, \lemmas^N)$ such that $\signature^N$ is
called the \emph{local signature} of $N$, $\axioms^N$ a set of \emph{local axioms} of $N$, and $\lemmas^N$ a set of \emph{local lemmas} of $N$. $\calL$ is a set of global definition links $\glinka{M}{\sigma}{N}$. Each link imports the mapped theory of $M$ (by the signature morphism $\sigma$)
as part of the theory of $N$. A node $N$ is globally reachable from a node $M$ via a signature morphism ${\sigma}$, ${\DG}\derives M \greaches {\sigma} N $ for short, iff
\begin{inparaenum}
\item either $M = N$ and $\presigma = id$, or
\item $\glinka{M}{{\sigma'}}{K} \in {\calL}$, and ${\DG}\derives K \greaches{\sigma''} N$, with $\presigma = {\sigma''} \circ {\sigma'}$.
\end{inparaenum}
The global signature (global axioms and global lemmata, respectively) of a node $N \in \calN$ is the union of its local signature (local axioms and local lemmata) and the mapped global signatures of all nodes from which $N$ is globally reachable.
A node is valid if all signature symbols occurring in its global axioms and lemmata are declared in its global signature. 
A development graph is well-defined, if all its nodes are valid. 

The \emph{maximal nodes} (root nodes) $\DGRoots{\DG}$ of a graph $\DG$ are all nodes without outgoing links.
$\Dom{\DG}(N) := \Signature{\DG}(N) \cup \Axioms{\DG}(N) \cup \Lemmas{\DG}(N)$ is the set of all signature symbols, 
axioms and lemmata visible in a node $N$. The \emph{local domain} of $N$, $\dom^{N} := \signature^N\cup\axioms^N \cup \lemmas^N$ is the
 set of all local signature symbols, axioms and lemmata of $N$.  
 The \emph{imported domain} $\Imports{\DG}(N)$ of $N$ in $\DG$ is the set of all signature symbols, axioms and lemmata imported 
 via incoming definition links.
  $\Dom{\DG} = \bigcup_{N \in \calN} \Dom{\DG}(N)$ is the set of all signature symbols,
  axioms and lemmata occurring in $\DG$. Analogously we define $\Signature{\DG}$,
  $\Axioms{\DG}$, and $\Lemmas{\DG}$.
	$\Dom{\DGRoots{\DG}} = \bigcup_{N \in \DGRoots{\DG}} \Dom{\DG}(N)$ is the set of all signature symbols, 
axioms and lemmata occurring in the maximal nodes of $\DG$.

Given a node $N \in {\cal N}$ its associated
  class $\Mod^{\DG}(N)$ of models (or $N$-models for short) consists of
  those $\Signature{\DG}(N)$-models $n$ for which
  \begin{inparaenum}[(i)]
  \item $n$ satisfies the local axioms $\axioms^N$, and
  \item for each $\glinka{K}{\presigma}{ N} \in {\cal
      S}$, $n\stroke_{\sigma}$ is a $K$-model.
\end{inparaenum}
In the following we denote the class of $\Sigma$-models 
that fulfill the $\Sigma$-sentences $\Psi$ by $\Mod_{\Sigma}(\Psi)$. 

Given a signature $\Sigma$ and $\AllAxioms, \AllLemmata \subseteq\Sen(\Sigma)$,
a \emph{support mapping $\Support$ for $\AllAxioms$ and $\AllLemmata$} assigns each 
lemma $\varphi\in\AllLemmata$ a subset $H\subseteq\AllAxioms\cup\AllLemmata$ such that
\begin{inparaenum}[(i)]
  \item $\Mod_{\SIGExtension{\symbols(H)\cup\symbols(\varphi)}{\Sigma}}(H)\models \varphi$ \footnote{
    where ${\SIGExtension{S}{\Sigma}}$ denotes the smallest valid sub-signature of $\Sigma$ containing $S$.
  }
  \item The relation $\supports\subseteq(\AllAxioms\cup\AllLemmata)\times\AllLemmata$ with
    $\Phi\supports\varphi \Leftrightarrow (
    \Phi\in\Support(\varphi) \vee \exists \psi. \Phi\in\Support(\psi) \wedge \psi\supports\varphi 
    )$
   is a well-founded strict partial order. 
  \end{inparaenum}
If $\DG$ is a development graph, then a support mapping $\Support$ is a \emph{support mapping for $\DG$} iff 
for all $N\in\DG$ $\Support$ is a support mapping for $\Axioms{\DG}(N)$ and $\Lemmas{\DG}(N)$.

We will now formalize the
requirements on development graphs that reflect our intuition of 
an appropriate structuring for formal specifications in the following principles.

The first principle is \emph{semantic appropriateness}, saying  that the structure of the development graph
should be a syntactical reflection of the relations between the various concepts in our
specification. This means that different basic specifications are located in different nodes
of the graph and the links of the graph reflect the logical relations between these
specifications.  
The second principle is \emph{closure} saying, for instance, that deduced knowledge 
should be located close to the axioms guaranteeing the proofs. Also the specification
defined by the theory of an individual node of a development graph should 
have a meaning of its own and provide some source of deduced knowledge.
The third principle is \emph{minimality} saying that each concept (or part of it) is only represented 
once in the graph. When splitting a monolithic theory into different theories
common foundations for these theories should be (syntactically) shared 
between them by being located at a unique node of the graph. 

We now translate these principles into syntactical criteria on
development graphs and into procedures of how to transform or refactor development
graphs. In a first step we formalize technical requirements to enforce the minimality-principle
in terms of development graphs. Technically, we demand that each signature symbol,
each axiom and each lemma has a unique location in the development graph. When we
enrich a development graph with more structure we forbid to have multiple copies of the same definition in different nodes. 
We therefore require that we can
identify for a given signature entry, axiom or lemma a \emph{minimal
  theory} in a development graph and that this minimal theory is
unique. We define:
\begin{definition}[Providing Nodes]
Let $\langle\calN,\calL\rangle$ be a development graph.
An entity $e$ is \emph{provided} in $N \in \calN$ iff $e \in \Dom{\langle\calN,\calL\rangle}(N)$ and
$\forall \glinka{M}{\presigma}{N}.\; e \not\in \Dom{\langle\calN,\calL\rangle}(M)$. Furthermore, 
\begin{compactenum}
\item
$e$ is \emph{locally} provided in $N$ iff additionally $e \in \dom^N$ holds.\vspace*{-1ex}
\item $e$ is provided \emph{by a link} $l : \glinka{M}{\presigma}{N}$
  iff $e$ is not locally provide in $N$ and $\exists e'\in \Dom{\langle\calN,\calL\rangle}(M).\; \sigma(e') = e$
  holds. In this case we say that $l$ provides $e$ from $e'$. $e$ is
  \emph{exclusively} provided by $l$ iff $e$ is not provided by any
  other link $l' \in \calL$.
\end{compactenum}
% We denote the set of elements exclusively provided by some link $l$ by $\providedby(l)\subseteq\Dom{\langle\calN,\calL\rangle }(N)$. 
\end{definition}
The closure-principle demands that there are no 
spurious nodes in the graph not contributing anything new. We combine
these requirements into the notion of location mappings:
\begin{definition}[Location Mappings]
  \label{def:location}
  Let $\DG = \langle \calN, \calL \rangle$ be a development graph.
  A mapping $\loc_{\DG} : \Dom{\DG} \to \calN$ is a \emph{location} mapping for $\DG$ iff
  \begin{compactenum}
  \item $\loc_{\DG}$ is surjective (closure)
  \item $\forall N \in \calN.\; \forall e \in \dom^N.\; loc_{\DG}(e) = N$ 
  \item $\forall e \in \Dom{\DG}.\; \loc_{\DG}(e)$ is the only node providing $e$ (minimality) 
  \end{compactenum}
 For a given $\loc_{\DG}$ we define $\loc_{\DG}^{-1}:\calN\to 2^{\Dom{\DG}}$ by
 \begin{compactitem}
  \item[] $\loc_{\DG}^{-1}(N) := \{ e\in\Dom{\DG} | \loc_{\DG}(e) = N \}$.
 \end{compactitem}
  We write $\loc$ and $\loc^{-1}$ instead of $\loc_{\DG}$ and $\loc^{-1}_{\DG}$ if $\DG$ is clear from the context.
\end{definition}

Based on the notion of location mappings we formalize our intuition of a 
\emph{structuring}. The idea is that the notion of being a structuring constitutes 
the invariant of the structure formation process and guarantees both, requirements 
imposed by the minimality-principle as well as basic conditions on a development graph
to reflect a given formal specification.

\begin{definition}[Structuring]\label{Ax:Structuring}
  Let ${\DG}=\langle {\cal N}, {\cal L} \rangle$ be a valid development graph, $\loc : \Dom{\DG} \to \calN$, 
  $\Sigma\in|\Sign|$, $\AllAxioms, \AllLemmata \subseteq \Sen(\Sigma)$ and $\Support$ be a support mapping for $\DG$.
  Then $(\DG, \loc, \Support)$ is a \emph{structuring} of $(\Sigma, \AllAxioms, \AllLemmata)$ iff
  \begin{compactenum}
  \item $\loc$ is a location mapping for $\DG$.
  \item \label{prop:preserve-elements-in-maximal-nodes}
	 let $\Dom{\DGRoots{\DG}} = \Sigma' \cup \AllAxioms' \cup \AllLemmata'$ then
	$\Sigma = \Sigma'$, $\AllAxioms = \AllAxioms'$ and $\AllLemmata \subseteq \AllLemmata'$.\vspace*{-1ex}
  \item \label{prop:support-relation-respected} $\forall \phi\in\Lemmas{\DG}\;.\; \forall \psi\in\Support(\phi).\; \exists \presigma.\;\loc(\psi) {\greaches{\presigma}} \loc(\phi) \wedge \presigma(\psi) = \psi$
  \end{compactenum}
\end{definition}

\section{Refactoring Rules}

In the following we present the transformation rules on 
development graphs that transform a structuring again into a structuring.
Using these rules we are able to structure the initially trivial development
graph consisting of exactly one node that comprises all given concepts step 
by step. This initial development graph consisting of exactly one node satisfies
the condition of a structuring provided that we have an appropriate support
mapping at hand. 

We define four types of structuring-invariant transformations:
\begin{inparaenum}[(i)]
  \item horizontal splitting and merging of development graph nodes, 
  \item vertical splitting and merging of development graph nodes, 
  \item factorization and multiplication of development graph nodes, and 
  \item removal and insertion of specific links. 
\end{inparaenum}
Splitting and merging as well as factorization and multiplication are dual operations. 
For lack of space and because we are mainly interested
in rules increasing the structure of a development graph we will omit the formal
specification of the merging and multiplication rules here.

\paragraph{Horizontal Split.} The first refactoring rule aims at the separation of specifications
in independent theories. In terms of the development graph a node is replaced
by a series of independent nodes; each of them contains a distinct part from a partitioning of the 
specification of the original node. In order to ensure a valid new development graph, 
each of the new nodes imports the same theories as the old node and contributes to the same theories
as the old node did. To formalize this rule we need constraints on how to
split a specification in different chunks such that local lemmata are always located in
a node which provides also the necessary axioms and lemmata to prove it. 
\begin{definition}
Let $\cS = (\DG, \loc, \Support)$ be a structuring of $(\Sigma, \AllAxioms, \AllLemmata)$
and $N \in \calN_{\DG}$. 
A partitioning ${\cal P}$ for $N$ is a set $\{N_1, \dots, N_k\}$ with $k > 1$ such that 
\begin{inparaenum}
\item $\signature^N  = \signature^{N_1} \uplus \ldots \uplus \signature^{N_k}$, 
      $\axioms^N  = \axioms^{N_1} \uplus \ldots \uplus \axioms^{N_k}$,
      $\lemmas^N  = \lemmas^{N_1} \uplus \ldots \uplus \lemmas^{N_k}$
\item $\signature^{N_i} \cup \axioms^{N_i} \cup \lemmas^{N_i} \not= \emptyset$ for $i = 1,\ldots, k$.
\end{inparaenum}
A node $N_i \in {\cal P}$ is \emph{lemma independent} iff 
$\Support(\psi) \cap (\axioms^{N} \cup \lemmas^{N}) \subseteq (\axioms^{N_i} \cup \lemmas^{N_i})$ for
all $\psi \in \lemmas^{N_i}$.
\end{definition}

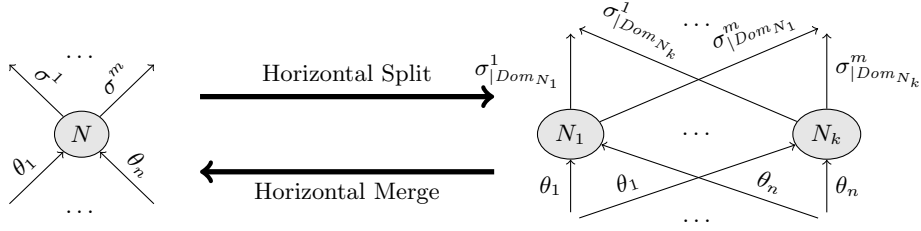
\begin{figure}[htb]
  \centering 
  \begin{tikzpicture}[node distance=7mm and 7mm]
    \node [namednode] (n) { $N$ }; 
    \node [above left=of n] (e1) {};
    \node [above right=of n] (e21) {};
    \node [below left=of n] (i1) {};
    \node [below right=of n] (i2) {};
    \draw[arrows=->] (i1) -- node[above,sloped] {$\pretheta_1$} (n);
    \draw[arrows=->] (i2) -- node[above,sloped] {$\pretheta_n$} (n);
    \draw[draw=none] (i1) -- node[sloped] {\dots} (i2);
    \draw[arrows=->] (n) -- node[above,sloped] {$\presigma^1$} (e1);
    \draw[arrows=->] (n) -- node[above,sloped] {$\presigma^m$} (e21);
    \draw[draw=none] (e1) -- node[sloped] {\dots} (e21);
    \begin{scope}[xshift=6.5cm]
    \node [namednode] (n1) { $N_1$ }; 
    \node [namednode,right=25mm of n1] (n2) { $N_k$ }; 
    \node [above=10mm of n1] (e1) {};
    \node [above=10mm of n2] (e2) {};
    \node [below=of n1] (i1) {};
    \node [below=of n2] (i2) {};
    \draw[arrows=->] (i1) -- node[left] {$\pretheta_1$} (n1);
    \draw[arrows=->] (i2) -- node[above,sloped,near start] {$\pretheta_n$} (n1);
    \draw[arrows=->] (i1) -- node[above,sloped,near start] {$\pretheta_1$} (n2);
    \draw[arrows=->] (i2) -- node[right] {$\pretheta_n$} (n2);
    \draw[draw=none] (i1) -- node[sloped] {\dots} (i2);
    \draw[arrows=->] (n1) -- node[left] {$\presigma^1_{|\Dom{N_1}}$} (e1);
    \draw[arrows=->] (n1) -- node[above,sloped,near end] {$\presigma^m_{|\Dom{N_1}}$} (e2);
    \draw[bend right=45,arrows=->] (n2) -- node[above,sloped,near end] {$\presigma^1_{|\Dom{N_k}}$} (e1);
    \draw[arrows=->] (n2) -- node[right] {$\presigma^m_{|\Dom{N_k}}$} (e2);
    \draw[draw=none] (e1) -- node[sloped] {\dots} (e2);
	  \draw[draw=none] (n1) -- node[sloped] {\dots} (n2);
  \end{scope}
  \draw[line width=2pt,arrows=->] ($ (n.east-|e21)!.1!(n1.west) + (0,5mm) $) -- 
        node[above,sloped]{Horizontal Split} ($ (n.east)!.9!(n1.west) + (0,5mm) $);
  \draw[line width=2pt,arrows=<-] ($ (n.east-|e21)!.1!(n1.west) - (0,5mm) $) -- 
        node[below,sloped]{Horizontal Merge} ($ (n.east)!.9!(n1.west) - (0,5mm) $);
\end{tikzpicture}
\caption{Horizontal Split and Merge}
\end{figure}

\begin{definition}[Horizontal Split]
	Let $\cS = (\langle\calN,\calL\rangle, \loc, \Support)$ be a
  \emph{structuring} of $(\Sigma, \AllAxioms,\AllLemmata)$, 
	${\cal P} = \{N_1, \ldots, N_k\}$ be a partitioning for some node $N \in \calN$ such that 
	each $N_i \in {\cal P}$ is lemma independent and $\loc^{-1}(N) = \dom^N$.
	The \emph{horizontal split} of $\cS$ wrt.\ $N$ and ${\cal P}$ 
	is $\cS' =  (\DG', \loc', \Support)$ with $\DG' = \langle\calN',\calL' \rangle$ where 
  \begin{compactenum}
	 \item $\calN' :=  \{N_1,\ldots, N_k\}\uplus (\calN\setminus N)$
   \item $\calL' :=  \{\glinka{M}{\presigma}{M'} \in {\cal L} | M \not= N \wedge M' \not= N \}$ \\
      \phantom{$\calL' :=$} $\cup \; \{\glinka{M}{\pretheta}{N_i} |  \glinka{M}{\pretheta}{N} \in {\cal L}, i \in \{1,\ldots,k \}\}$ \\
      \phantom{$\calL' :=$} $\cup \; \{\glinka{N_i}{\pretau_{|\Dom{N_i}}}{M} | \glinka{N}{\pretau}{M} \in {\cal L}, i\in\{1,\ldots, k\} \} $
   \item $\loc'(e) :=  N_i$  if $e\in\dom^{N_i}$ for some $i\in\{1,\ldots, k\}$ and $\loc'(e) :=\loc(e)$ otherwise.
 \end{compactenum}
 such that $\Signature{\DG'}(N_i)$ are valid signatures and $\axioms_i, \lemmas_i\subseteq\Sen(\Signature{\DG'}(N_i))$ for $i=1,\ldots,k$.
\end{definition}

\paragraph{Vertical Split.} Similar to a horizontal split we introduce a vertical split which divides a node
into two nodes and locates one node on top of the other. While all outgoing links start
at the top node, we are free to reallocate incoming links to either node. 
\begin{definition}[Vertical Split] 
  Let $\cS = (\langle\calN,\calL\rangle, \loc, \Support)$ be a 
	\emph{structuring} of $(\Sigma, \AllAxioms, \AllLemmata)$ and
	${\cal P} = \{N_1, N_2\}$ be a partitioning for some $N\in\calN$ such that
	$N_1$ is lemma independent. 
  Then, the vertical split $\cS$ wrt. $N$ and $\cal P$ is 
	$\cS' = (\DG',\loc', \Support)$ with $\DG' = \langle\calN',\calL'\rangle$ where
  \begin{align*}
    \calN' := & \{N_1,N_2\}\uplus (\calN\setminus N)\\
    \calL' := & \{\glinka{M}{\presigma}{M'} \in {\cal L} | M \not= N \wedge M' \not= N \} \cup \{ \glinka{N_1}{id}{N_2} \} \\
    & \cup \{ \glinka{M}{\presigma}{N_1} \mid  \glinka{M}{\presigma}{N}\in\calL\}
      \cup \{ \glinka{N_2}{\presigma}{M} \mid  \glinka{N}{\presigma}{M}\in\calL\}\\
    \loc'(e) = &\left\{
      \begin{array}{ll}
        N_2 & \text{if } \loc(e) = N\text{ and } e\in\Dom{\DG'}(N_2)\\
        N_1 & \text{if } \loc(e) = N\text{ and } e\not\in\Dom{\DG'}(N_2)\\
        \loc(e) & \text{otherwise}
      \end{array}
    \right.
  \end{align*}
  such that $\Signature{\DG'}(N_i), i=1,2$, are valid signatures and
  $\axioms_i, \lemmas_i\subseteq\Sen(\Signature{\DG'}(N_i))$, $i=1,2$. Conversely,
  \emph{$\cS$ is a vertical merge of $N_1$ and $N_2$ in $\cS'$}.
\end{definition}

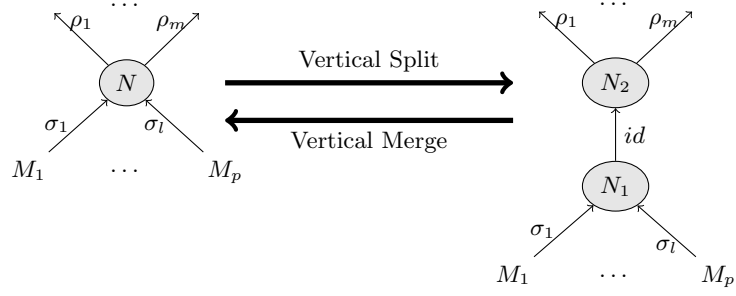
\begin{figure}[tb]
  \centering 
  \begin{tikzpicture}[node distance=7mm and 7mm]
    \node [namednode] (n) { $N$ }; 
    \node [above left=of n] (o1) {};
    \node [above right=of n] (om) {};
    \node [below left=of n] (m1) {$M_1$};
    \node [below right=of n] (mp) {$M_p$};
    \draw[arrows=->] (m1) -- node[left] {$\presigma_1$} (n);
    \draw[arrows=->] (mp) -- node[left] {$\presigma_l$} (n);	
    \draw[draw=none] (o1) -- node[sloped] {\dots} (om);
    \draw[arrows=->] (n) -- node[above] {$\rho_1$} (o1);
    \draw[arrows=->] (n) -- node[above] {$\rho_m$} (om);
    \draw[draw=none] (m1) -- node[sloped] {\dots} (mp);
    \begin{scope}[xshift=6.5cm]
    \node [namednode] (n2) { $N_2$ }; 
    \node [above left=of n2] (o1) {};
    \node [above right=of n2] (om) {};
    \node [namednode,below=of n2] (n1) {$N_1$};
    \node [below left=of n1] (mm1) {$M_1$};
    \node [below right=of n1] (mmp) {$M_p$};
    \draw[arrows=->] (n2) -- node[above] {$\rho_1$} (o1);
		\draw[arrows=->] (n2) -- node[above] {$\rho_m$} (om);
		\draw[draw=none] (o1) -- node[sloped] {\dots} (om);
    \draw[arrows=->] (n1) -- node[right] {$id$} (n2);
    \draw[arrows=->] (mm1) -- node[left] {$\presigma_1$} (n1);
    \draw[arrows=->] (mmp) -- node[below] {$\presigma_l$} (n1);
    \draw[draw=none] (mm1) -- node[sloped] {\dots} (mmp);
    \node at ($(n2.west)!.5!(n1.west-|n2.west)$) (center) {};
  \end{scope}
  \draw[line width=2pt,arrows=->] ($ (n2.east-|mp)!.0!(center) $) -- 
        node[above,sloped]{Vertical Split} ($ (n2.west-|mm1)!.0!(center) $);
  \draw[line width=2pt,arrows=<-] ($ (n2.east-|mp)!.0!(center) - (0,5mm) $) -- 
        node[below,sloped]{Vertical Merge} ($ (n2.west-|mm1)!.0!(center) - (0,5mm) $);
\end{tikzpicture}
\caption{Vertical Split and Merge}
\end{figure}

\newcommand{\AxiomAssoc}[1]{\ensuremath{\forall x,y,z \,.\, x \operatorname{#1} (y \operatorname{#1} z) = (x \operatorname{#1} y) \operatorname{#1} z}}
\newcommand{\AxiomComm}[1]{\ensuremath{\forall x,y \,.\, x \operatorname{#1} y = y \operatorname{#1} x}}
\newcommand{\AxiomId}[2]{\ensuremath{\forall x \,.\, x \operatorname{#1} #2 = x}}
\newcommand{\AxiomInv}[3]{\ensuremath{\forall x \,.\, x \operatorname{#1} \operatorname{#2}(x) = #3}}

\begin{example}\label{ex:example1}
  We illustrate the horizontal and vertical split rules by considering a single theory axiomatizing a Field with binary operations $+$ and $\times$ consisting of a Distributivity axiom ($\Phi_{D}:=\forall x,y,z . x\times(y + z) = x\times y + x\times z$) and
  the axioms of an Abelian Group for $+$ and $\times$, respectively ($\Phi_{AG}^+:=\AxiomAssoc{+}, \AxiomComm{+}, \AxiomId{+}{0}, \AxiomInv{+}{-}{0}$ and $\Phi_{AG}^{\times}:=\AxiomAssoc{\times}, \AxiomComm{\times}, \AxiomId{\times}{1}, \AxiomInv{\times}{inv}{1}$). Assume axioms are contained in a single node \textsf{Field}, which forms a trivial structuring. In a first step we can split that node vertically by separating the distributivity axiom from the other axioms. In a second step we can separate the Abelian Group axioms for $+$ and $\times$ by a horizontal split. This is shown in the following Figure:

\begin{center}
\begin{tikzpicture}[node distance=7mm and 7mm]
    \node [namednode] (field) { $\begin{array}{c}\Phi_D\\\Phi_{AG}^{+}\\ \Phi_{AG}^{\times}\end{array}$ }; 
    \node [below=.6cm of field] (version1) {(0)};
    \node [right=3.3cm of version1] (version2) {(1)};
    \draw[line width=2pt,arrows=->] (version1) -- node[above,sloped]{Vertical Split} (version2);
    \node [right=4.3cm of version2] (version3) {(2)};
    \draw[line width=2pt,arrows=->] (version2) -- node[above,sloped]{Horizontal Split} (version3);
    \begin{scope}[xshift=4cm,yshift=.8cm]
      \node [namednode] (distr) {  $\Phi_D$ }; 
      \node [namednode,below=of distr] (abelgroups) { $\Phi_{AG}^{+}, \Phi_{AG}^{\times}$ };
      \draw[arrows=->] (abelgroups) -- node[right] {$id$} (distr);
    \end{scope}
    \begin{scope}[xshift=9cm,yshift=.8cm]
      \node [namednode] (distr2) { $\Phi_D$  }; 
      \node [namednode,below left=of distr2] (abelgroupplus) { $\Phi_{AG}^{+}$};
      \node [namednode,below right=of distr2] (abelgrouptimes) { $\Phi_{AG}^{\times}$};
      \draw[arrows=->] (abelgroupplus) -- node[left] {$id$} (distr2);
      \draw[arrows=->] (abelgrouptimes) -- node[right] {$id$} (distr2);
    \end{scope}
  \end{tikzpicture}
\end{center}
\end{example}

\paragraph{Factorization.}
The factorization rule allows one to merge equivalent specifications into a single generalized specification and then to 
represent the individual ones as instantiations of the generalized specification. A precondition of this rule is that
all individual specifications inherit the same (underlying) theories. 
\begin{definition}[Factorization]
  Let $\cS = (\langle\calN,\calL\rangle, \loc, \Support)$ be a
  \emph{structuring} of $(\Sigma, \AllAxioms, \AllLemmata)$.
	Let $K_1, \ldots, K_n, M_1, \ldots, M_p \in \calN$ with $p > 1$ such that
	$\signature^{M_j} \cup \axioms^{M_j} \not= \emptyset$ and
  $\exists \presigma_{i,j}. \;\glinka{K_i}{\presigma_{i,j}}{M_j} \in \calL$ for
	$i = 1,\ldots, n, j = 1,\ldots, p$. 
	
	Suppose there are sets $\signature$, $\axioms$ and $\lemmas$ with 
	$(\signature \cup \axioms \cup \lemmas) \cap \Dom{\DG} = \emptyset$ and signature morphisms $\theta_1,\ldots, \theta_p$ and
	$\sigma_1,\ldots,\sigma_n$ such that 
	\begin{compactitem}[-]
	\item $\forall e \in \Dom{\DG}(K_i). \; \theta_j(\sigma_i(e)) = \sigma_{i,j}(e)$ and $\sigma_{i,j}(e) = e \vee \sigma_{i,j}(e) \not\in \Dom{\DG}$
	\item $\signature^{M_j} \subseteq \theta_j(\signature) \subseteq \Dom{\DG}(M_j)$, $\axioms^{M_j} \subseteq \theta_j(\axioms) \subseteq \Dom{\DG}(M_j)$
  \item $\forall e \in \lemmas$ holds $\exists l \in \{1,\ldots p\}. \; \theta_l(e) \in \lemmas^{M_l}$,
	 $\theta_i(e) = \theta_j(e)$ implies $i = j$ and $\theta_j(e) \in \Dom{\DG}$ implies $\loc(\theta_j(e)) \in M_j$
	\item there is a support mapping 
	$\Support_N$ for $\axioms \cup \bigcup_{i=1,...,n} \sigma_i(\Dom{\DG}(K_i))$ and $\lemmas$.
	\end{compactitem}
	Then $\cS' = (\langle\calN',\calL'\rangle, \loc', \Support')$ is a factorization of $\cS$ wrt. $M_1, \ldots$, $M_p$ and $\Support_N$ iff
	\begin{align*}
	  \calN' := & \{N\} \cup \{N_j | j\in\{1,\ldots p\} \} \cup \calN \setminus\{M_1,\ldots M_p\}\\
							& \text{ with } N = \langle \signature, \axioms, \lemmas \rangle, N_j = \langle \emptyset, \emptyset, \lemmas^{M_j} \setminus \theta_j(\lemmas) \rangle \\
    \calL' := & \{\glinka{K}{\presigma}{K'} \in {\cal L} | K, K'\not\in \{M_1,\ldots M_p \} \\
		          & \cup \{\glinka{K_i}{\presigma_i}{N} | \glinka{K_i}{\presigma_{i,j}}{M_j}, j\in\{1,\ldots p\}, i\in\{1,\ldots n\} \} \\
							& \cup \{\glinka{N}{\theta_j}{N_j} | j\in\{1,\ldots p\} \} \\
							& \cup \{\glinka{K}{\pretau}{N_j} | \glinka{K}{\pretau}{M_j} \wedge (\forall i\in\{1,\ldots n\}. 
							          K \not= K_i \wedge \pretau \not= \presigma_{i,j}) \\
							& \cup \{\glinka{N_j}{\pretau}{K} | \glinka{M_j}{\pretau}{K} \in \calL, j\in\{1,\ldots p\} \} \\
		\loc'(x) := &\left\{
                \begin{array}{ll}
                   N    & \text{if } x \in \Dom{\DG'}(N) \setminus \bigcup_{i=1,\ldots,n} \Dom{\DG'}(K_i) \\
                   N_j  & \text{if } x \in \Dom{\DG}(N_j)  \text{ and } \forall \glinka{K}{\presigma}{N_j}. \, x \not\in \Dom{\DG'}(K)  \\
									 \loc(x) & \text{otherwise.}
                \end{array}
               \right.\\
		\Support' := & \Support \cup \Support_N.
	\end{align*}
\end{definition}

\begin{figure}[tb]
  \centering 
  \begin{tikzpicture}[node distance=7mm and 7mm]
	  \node [] (dummy) {};
    \node [namednode, above=of dummy] (n) { $M_1$ };
		\node [below left=of n] (f11) {};
    \node [above left=of n] (e11) {};
    \node [above=of n] (e12) {};
    \node [namednode,right=of n] (m) { $M_p$ }; 
    \node [above=of m] (e21) {};
		\node [below right=of m] (f1p) {};
    \node [above right=of m] (e22) {};
		\draw[draw=none] (n) -- node[sloped] {\dots} (m);

    \node [namednode,below=of dummy] (i1) { $K_1$ };
    \node [namednode,right=of i1] (i2) { $K_n$ };
    \draw[arrows=->] (i1) -- node[above,midway,sloped] {$\presigma_{1,1}$} (n);
    \draw[arrows=->] (i2) -- node[above,near start,sloped] {$\presigma_{n,1}$} (n);
    \draw[arrows=->] (i1) -- node[above,near start,sloped] {$\presigma_{1,p}$} (m);
    \draw[arrows=->] (i2) -- node[below,midway,sloped] {$\presigma_{n,p}$} (m);
    \draw[draw=none] (i1) -- node[sloped] {\dots} (i2);
    \draw[arrows=->] (n) -- node[above,sloped] {} (e11);
    \draw[arrows=->] (n) -- node[below,sloped] {} (e12);
		\draw[arrows=->] (f11) -- node[above,sloped] {} (n);
    \draw[draw=none] (e11) -- node[sloped] {\dots} (e12);
    \draw[arrows=->] (m) -- node[above,sloped] {} (e21);
    \draw[arrows=->] (m) -- node[above,sloped] {} (e22);
		\draw[arrows=->] (f1p) -- node[above,sloped] {} (m);
    \draw[draw=none] (e21) -- node[sloped] {\dots} (e22);
    \node at (dummy.east-|e22) (center1) {};
    \begin{scope}[xshift=8.5cm,yshift=0mm]
		\node [namednode] (n0) { $N$ };
		\node [namednode, below left= of n0] (k1) { $K_1$ };
		\node [namednode, below right= of n0] (kn) { $K_n$ };
		\node [namednode, above left=of n0] (n1) { $N_1$ };
		\node [namednode, above right=of n0] (n2) { $N_p$ };
		\node [below left=of n1] (f21) {};
		\node [above left=of n1] (e11) {};
    \node [above=of n1] (e12) {};
		\node [above right=of n2] (e21) {};
		\node [below right=of n2] (f2p) {};
    \node [above=of n2] (e22) {};
		\node [below= of n1] (f1) { };

    \draw[draw=none] (n1) -- node[sloped] {\dots} (n2);
    \draw [arrows=->] (n0) -- node[above,sloped] {$\pretheta_1$} (n1);
		\draw [arrows=->] (n0) -- node[above,sloped] {$\pretheta_p$} (n2);
    \draw[arrows=->] (k1) -- node[above,near start,sloped] {$\presigma_1$} (n0);
    \draw[arrows=->] (kn) -- node[above,midway,sloped] {$\presigma_n$} (n0);
    \draw[draw=none] (k1) -- node[sloped] {\dots} (kn);
		\draw[arrows=->] (n1) -- node[above,sloped] {} (e11);
    \draw[arrows=->] (n1) -- node[below,sloped] {} (e12);
		\draw[arrows=->] (f21) -- node[above,sloped] {} (n1);
		\draw[arrows=->] (f2p) -- node[above,sloped] {} (n2);
		\draw[arrows=->] (n2) -- node[above,sloped] {} (e21);
    \draw[arrows=->] (n2) -- node[below,sloped] {} (e22);
		\draw[draw=none] (e11) -- node[sloped] {\dots} (e12);
		\draw[draw=none] (e21) -- node[sloped] {\dots} (e22);
    \node at (n0.west-|n1) (center2) {};
  \end{scope}
  %\draw[line width=2pt,arrows=->] ($ (center1)!.1!(center2)  $) -- 
  %      node[above,sloped]{Factorization} ($ (center1)!.9!(center2)  $);
	\draw[line width=2pt,arrows=->] ($ (center1)!.1!(center2) - (0,1mm) $) -- 
        node[above,sloped]{Factorization} ($ (center1)!.9!(center2) - (0,1mm)$);
  %\draw[line width=2pt,arrows=<-] ($ (center1)!.1!(center2) - (0,5mm) $) -- 
  %      node[below,sloped]{Multiplication} ($ (center1)!.9!(center2) - (0,5mm) $);
\end{tikzpicture}
\caption{Factorization (with $\presigma_{i,j} := \pretheta_j \circ \presigma_i$)}
\end{figure}
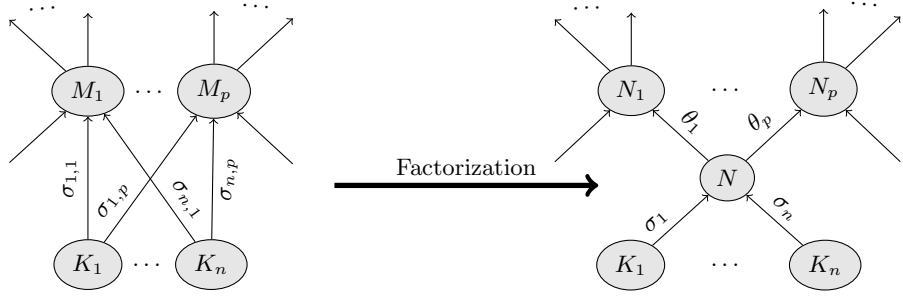
	
\begin{example}\label{ex:example2}
  Consider again our example a Field axioms, which we have transformed into the structuring (3) (p.~\pageref{ex:example1}). 
  On the last structuring (3) we can apply the factorization rule to extract the general abelian group axioms ($\Phi_{AG}^{\circ}:=\AxiomAssoc{\circ}, \AxiomComm{\circ}, \AxiomId{\circ}{e}, \AxiomInv{\circ}{i}{e}$) 
and obtain the respective axioms for $+$ and $\times$ by morphisms
\begin{math}
  \sigma_1 := \circ \mapsto +, e \mapsto 0, i \mapsto -
\end{math}
and $\sigma_2 := \circ \mapsto \times, e \mapsto 1, i \mapsto \operatorname{inv}$. 
This is illustrated in the following diagram and the final structuring contains 5 axioms and the initial structuring contained 9 axioms. 
\begin{center}
\begin{tikzpicture}[node distance=7mm and 7mm]
  \node [namednode] (distr2) { $\Phi_D$  }; 
  \node [namednode,below left=of distr2] (abelgroupplus) { $\Phi_{AG}^{+}$};
  \node [namednode,below right=of distr2] (abelgrouptimes) { $\Phi_{AG}^{\times}$};
  \draw[arrows=->] (abelgroupplus) -- node[left] {$id$} (distr2);
  \draw[arrows=->] (abelgrouptimes) -- node[right] {$id$} (distr2);
  \begin{scope}[xshift=6cm,yshift=0cm]
    \node [namednode] (distr3) { $\Phi_D$  }; 
    \node [namednode,below left=.7cm of distr3] (abelgroupplus) { $\emptyset$ };
    \node [namednode,below right=.7cm of distr3] (abelgrouptimes) { $\emptyset$};
    \draw[arrows=->] (abelgroupplus) -- node[left] {$id$} (distr3);
    \draw[arrows=->] (abelgrouptimes) -- node[right] {$id$} (distr3);
    \node [namednode,below=1.2cm of distr3] (abelgroup) { $\Phi_{AG}^{\circ}$  }; 
    \draw[arrows=->] (abelgroup) -- node[above] {$\sigma_1$} (abelgroupplus);
    \draw[arrows=->] (abelgroup) -- node[above] {$\sigma_2$} (abelgrouptimes);
  \end{scope}
    \node [below=2cm of distr2] (version3) {(3)};
    \node [right=5.2cm of version3] (version4) {(4)};
    \draw[line width=2pt,arrows=->] (version3) -- node[above,sloped]{Factorization} (version4);
  \end{tikzpicture}
\end{center}
\end{example}

The factorization rule only covers a sufficient criterion
demanding that each theory imported by a definition link to one specification is also imported via definition links by all
other specifications. The more complex case in which a theory is imported via a path of links can be handled by
allowing one to shortcut a path in a single global link. This results in the following rule.
			
\begin{definition}[Transitive Enrichment]
Let $\cS = (\langle \calN, \calL \rangle, \loc, \Support)$ 
be a structuring of $(\Sigma, \AllAxioms, \AllLemmata)$,
$K, N \in \calN$ and there is a path $K \greaches{\presigma}{N}$ between both. Then,
$\cS' = (\langle \calN, \calL \cup \{\glinka{K}{\presigma}{N}\} \rangle, \loc, \Support)$
is a transitive enrichment of $\DG$.
\end{definition}

Definition links in a development graph can be redundant, if
there are alternatives paths which have the same morphisms or if they are not used in any
reachable node of the target. We formalize these notions as follows:

\begin{definition}[Removable Link]\label{def:remlink}
  Let $\cS = (\DG, \loc, \Support)$ ($\DG = \langle \calN, \calL \rangle$) be a structuring of $(\Sigma, \AllAxioms, \AllLemmata)$.
	Let $l \in \calL$ and $\DG' = \langle \calN, \calL \setminus \{l\} \rangle$.
	$l$ is removable from $\cS$ and $\cS' = (\DG', \loc, \Support)$ is a \emph{reduction} of $\cS$ iff 
	\begin{compactenum}
	\item $\forall l': \glinka{M}{\presigma}{N}.$ if $l'$ 
          provides exclusively $\sigma(e)$ from some $e\in \Dom{\DG}(M)$ then $e \in \Dom{\DG'}(N)$ and $l \not= l'$;
	\item $\forall e\in\Dom{\DG}. \forall M\in \DGRoots{\DG}.$ if $\loc(e) \greaches{\presigma} M$
	 then there exists $M' \in \DGRoots{\DG'}$ such that \linebreak $\loc(e) \greaches{\presigma} M'$;
  \item $\forall \phi\in\Lemmas{\DG}. \; \Support(\phi) \subseteq \Dom{\DG'}(N)$ and
	      $\forall \Signature{\DG}^{loc}(N) \subseteq \Dom{\DG'}(N)$.
	\end{compactenum}
\end{definition}

\begin{theorem}[Structuring Preservation]\label{Thm:Soundness}
  Let $\cS := (\DG, \loc, \Support)$ ($\DG = \langle \calN, \calL \rangle$) be a structuring of $(\Sigma, \AllAxioms, \AllLemmata)$. Then
	\begin{compactenum}
	\item  every horizontal split of $\cS$ 
	wrt.\ some $N\in\calN$ and 
	partitioning $\cal P$ of $N$,
	\item  every vertical split of $\cS$ wrt.\ some $N\in\calN$ and 
	partitioning $\cal P$ of $N$,
	\item  every factorization of $\cS$ wrt.\ 
	nodes $M_1, \ldots M_p\in\calN$,
	\item  every transitive enrichment of $\cS$, and  
	\item  every reduction of $\cS$ 
	\end{compactenum}
	is a structuring of $(\Sigma, \AllAxioms, \AllLemmata)$.
\end{theorem}
The theorem follows from the soundness proofs for each rule  given in Appendix~\ref{sec:soundness proofs}.

\section{Refactoring Process}

In order to evaluate the refactoring rules on real theories we have
implemented the development graphs and the rules in Scala\footnote{\url{http://www.scala-lang.org/}} and added
support to read formulas in TSTP format~\cite{tstp} using the Java
parser from~\cite{tptpparser}. The support mapping is given as an
extra datastructure representing the information which formula has
been used in the proof of a theorem. In the case of TSTP we extract
that information from the files by using the names of the
formulas. Since the TSTP format does not include signature
declarations, we add declarations for all occurring symbols in a TSTP
file in an initialization step. We used the untyped part of TSTP and hence the declarations
only contain arity information but no types. 

The refactoring rules are parameterized over the theories and possibly
the subsets of the local signature, axioms and lemmata to split over.
To compute the parametric information we provided some basic heuristic
tactics. Using the support mapping, we define that an axiom
(resp. lemma) depends on a symbol declaration, if the symbol occurs in
the axiom (resp. lemma) and a lemma depends on another axiom or lemma,
if the latter is in its support mapping. A symbol declaration is
always independent. This dependency relation induces a partial order
on the local domain of each node in a development graph.

\paragraph{Tactic for horizontal split.} This rule requires the
partitioning of the local signature, axioms and lemmas for a given
theory into independent parts such that given the same imports than
the original node, each part is a valid theory and lemma independent
of the other part. We implemented a heuristic that given a local
domain of some node, searches for a largest subset which has a
non-empty intersection of its occurring symbols and supporting axioms
and lemmata.  If such a set exists, the largest such set is used to split the theory 
horizontally into that set and the rest.

\paragraph{Tactics for vertical split.} The rule requires to find a
subset of the local domain, which is independent of the rest and use
it as the content of the lower theory. We implemented two heuristics
to search for this subset. First, we consider all maximal elements
wrt. the dependency relation and use that as content for the new upper
theory constructed by vertical split. Second, we consider all minimal
elements and use it as content for the lower theory constructed by
vertical split. These two tactics allow one to incrementally split a theory into layered slices of the dependency relation. 

\paragraph{Tactic for factorization.} This rule requires to find
isomorphic subsets in two different theories to factorize over. The
notion of isomorphism between formulas is very strict, as we only
search for renamings. Furthermore, we extended the isomorphism to the
support mapping such that lemmata can only be identified with
isomorphic lemmata which supporting axioms and lemmata are also
isomorphic wrt. the same renaming. Thus, an axiom can never be
factorized with a lemma and vice-versa. Even with that strict notion,
computation of such subsets is already expensive. If the entire local
domain of a given node is isomorphic to the local domain of the second
node, both nodes are factorized according the definition of the
factorization rule. If the identified subset in the first node does
not cover the complete second node, we first try to split the second
node to isolate the subset. To this end we first try to split the
second node horizontally using the identified subset. If that fails,
we first try to split vertically using the subset for the upper part
and finally as the lower part. If one of these splittings was
successful, the factorization is applied on the isolated
part. Otherwise the factorization fails.

In addition to these main tactics, we have implemented the tactics to
delete superfluous links as well as deletion of empty nodes which
technically corresponds to vertically merging the empty node with
their importing theories. 

\paragraph{Automatic Procedure.} In order to automate the theory formation process
we have implemented the usual tacticals to describe more complex search behaviors. 
The tactic language is defined as follows starting from the basic tactics described above:

\[
\begin{array}{rcl}
  T & ::= & SplitHorizontal\, |\, SplitVerticallyMaximal \,|\, SplitVerticallyMinimal\\ 
  &|& Factorize\, |\, RemoveSuperfluousEmptyTheories\\
  &|& T* \,|\, T+ \,|\, T;T \,|\, T\ onfail\ T  
\end{array}
\]

The tactics take as argument a structuring and if they could be
applied, return a new structuring and otherwise fail. The tacticals
for as many as possible iteration ($*$), as many as possible but at
least one ($+$) and sequencing ($;$) are standard. The tactical
$onfail$ executes the second tactic expression only if the first
failed. Using this language we have implemented the following
automatic procedure.  The goal of the procedure is starting from an
unstructured graph, i.e. a single theory containing all declarations,
axioms and lemmata, to search for possibilities to factorize common
patterns. Factorization is only possible if at least one application
of the horizontal split rule was possible, which in turn may require
the application of a preparatory vertical split. Following that
initial part, we try to split further vertically using the maximal
elements of the theory and finally removing the superfluous links and
empty theories. Hence, the initial phase of the automation consists of
\[
\begin{array}{rcl}
inittac & \dot{\equiv} &  ((SplitVerticallyMinimalEntries+;SplitHorizontally*)\\
        && onfail\ SplitHorizontally+);\\
        && SplitVerticallyMaximalEntries*;\\
        && RemoveSuperfluousEmptyTheories*
\end{array}
\]
That initialization tactic succeeds only if at least one vertical split or one horizontal split 
could be done. Following that, we start to factorize. If at least one factorization was possible, 
we first clean up the structuring by removing  superfluous links and
empty theories before trying again to split vertically. The overall tactic is thus 
\[
inittac;\begin{array}[t]{l}
(Factorize+;RemoveSuperfluousEmptyTheories*;\\
SplitVerticallyMinimalEntries*)*
\end{array}
\]

\section{Evaluation}

We have applied the factorization procedure presented in the previous section
to TSTP versions of the Mizar library articles~\url{www.mizar.org},
which have been created by Joseph Urban and are available at
\url{http://www.cs.miami.edu/~tptp/MizarTPTP/TPTPArticles/}. This is a
collection of 922 files in TSTP format
(\url{www.cs.miami.edu/~tptp/TSTP}) where theorems are annotated by
information which theorems and axioms have been used in their proofs.
\begin{figure}[t]
  \centering
  \setlength\tabcolsep{1em}
  \begin{tabularx}{\linewidth}{lcccc}
  \textbf{Article} & \textbf{Axioms} & \textbf{Theorems} & \textbf{Reduction} & \textbf{Timeout} \\ 
  \input{results.tex}
  \end{tabularx}
\label{fig:results}
\caption{Factorization results on TSTP versions of the Mizar articles}
\end{figure}
The files consist of the axioms and theorems of each article including
all directly included articles, but without transitive expansion of
all inclusions.  Hence, the knowledge in each file is already quite
tailored to the knowledge necessary to define the additional
mathematical concepts and to enable the proofs of the theorems.  We
have run the procedure on all examples with a timeout of 5
minutes each. The environment was a virtual machine with 4 virtual CPUs,
16GB RAM, under openSuSE 12.2 64-bit, running on a host with 2 Intel
Xeon Westmere E5620 QuadCore CPUs, 2,4GHz, 96GB RAM and VMware ESXi
4.1.

\piccaption{Resulting DG \label{fig:dg-final}}

For most articles no factorization has been found. However,
there are 13 articles where factorization was possible, which are presented in
the table Fig.~\ref{fig:results}. 
The results are summarized in the following format: for each file we
indicate in the \textbf{Axioms} column the number of axioms in the
initial development graph and the final development
graph. Analogously, the \textbf{Theorems} column indicates the number
of theorems respectively in the initial and the final development
graph.  The \textbf{Reduction} column indicates how much the
factorization \ reduced the overall \ number  of \  axioms and theorems. The
last column indicates if the 
\parpic[r]{\includegraphics[width=.34\linewidth]{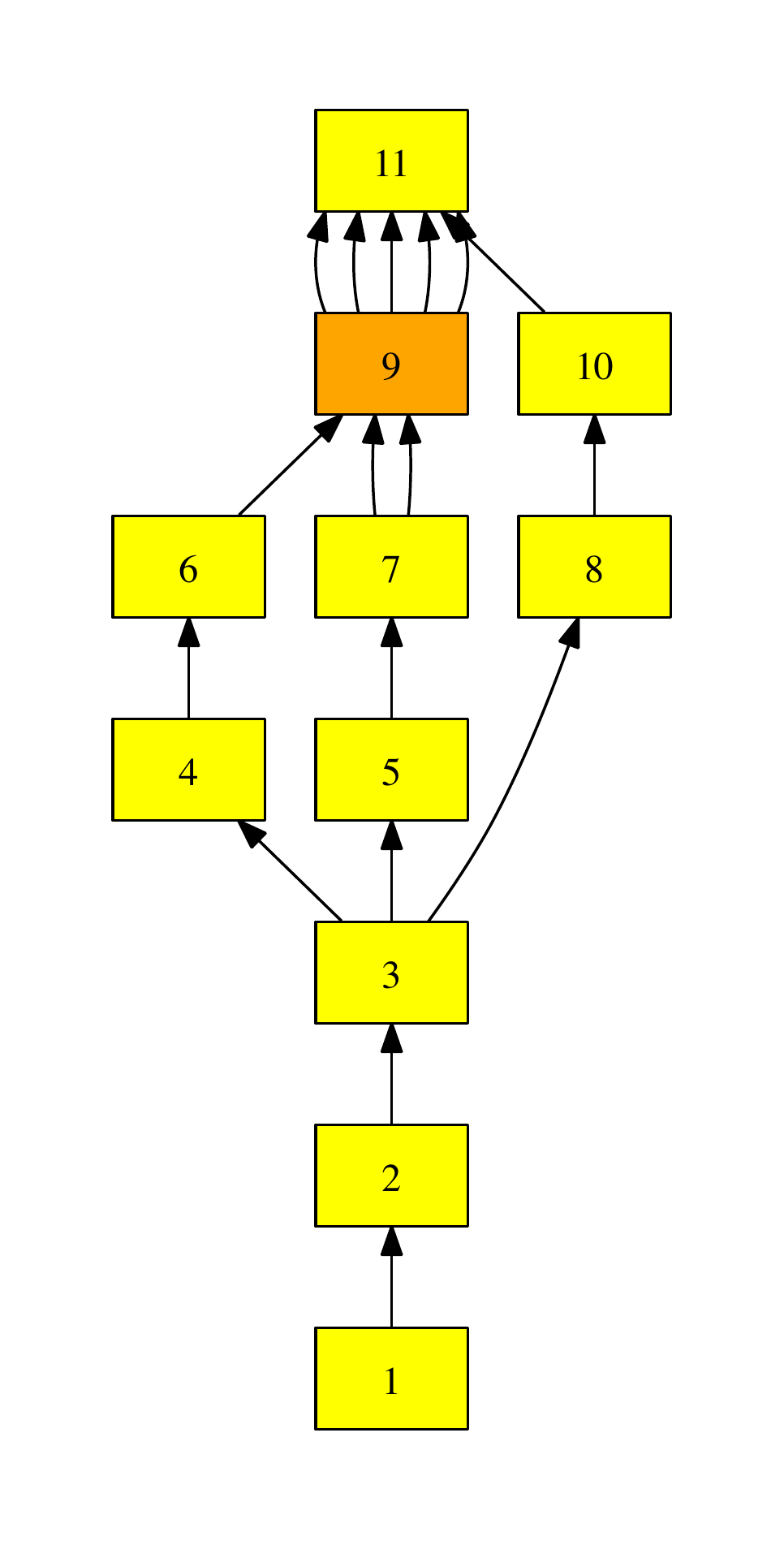}}
\noindent
automatic  procedure had terminated within
the 5 minutes time frame or timeout was reached.

While reducing the number of axioms by factorization is already
interesting in order to reduce the search space for automatic provers,
reducing the number of theorems is more interesting as it means less
theorems to prove.
For all but one file where factorizations have been found, only axiom
factorization have been found. However, in the article
\textbf{membered. top.rated} obtained from the Mizar
article~\cite{membered-03} ``On the Sets Inhabited by Numbers'' we
could factorize 36 theorems into 16 theorems. On closer inspection
this is not surprising because it concerned theorems about sets of
reals, sets of rationals, sets of integers, sets of naturals and sets
of complex numbers, all defined and proved according to the same
schema. The resulting development graph is shown on the right side of
Fig.~\ref{fig:results}, and the factor theory containing the 5
theorems, from which all others are obtained by renaming, is node
9 in gray/orange. The factorization is visible via the 5 outgoing edges
towards node 11 which are annotated with the respective morphisms.

\section{Related Work and Conclusion}

Related to the structuring of theories, there is a
large work on anti-unification, i.e. computing common generalizations of 
different formuala or theories (e.g. \cite{FP90,NK07,GK2014}). The resulting
structuring approach is primarily botton-up and driven by the pure existence
of anti-unifiers. In contrast, our approach is top-down as it introduces 
measures for the intended structuring (i.e. semantic appropriateness, closure 
and minimality) to guide the formation process. For example, we split up 
theories in smaller ones but that are still self-contained in the sense that 
each theorem of the original theory can be proven in one of the new (smaller) 
ones. Anti-unification is an important technique to test the applicability 
of the factorization rule, for instance, but applicability of a rule is not 
the driving force of the formation process.  

In this paper we were concerned with trying to reveal shared
definitions, axiomatizations and theorems in a given formal
theory. Based on \emph{structurings} which extend development graphs
with notions to exclude redundancies and include dependency
information, we presented a set of rules on structurings. We
implemented the rules with simple heuristics to detect isomorphic
subsets which are sufficient to find simple factorization and applied
it to the TSTP formulations of the Mizar articles. Not surprisingly,
not many factorizations could be found, which is due to Mizar's
non-transitive reuse principle of other articles and the fact that
these were chosen carefully by the authors of the Mizar
article. Moreover, the heuristics to compute isomorphic axioms and
theorems was very restricted. However, a few factorizations could be
found, and especially one were the number of theorems could be halved.
This indicates that adding theory morphisms to the Mizar language may
be useful, but that needs to be confirmed by further analysis of
larger subsets. On the other hand the non-transitive import mechanisms
of Mizar already seems to allow for a good organization of the
knowledge.  That kind of mechanism is typically not implemented in
specification languages, but exists in development graphs in form of
local axiom links.

Future work will consist of analyzing larger 
subsets of the whole Mizar library, i.e. sets of Mizar articles, for
possible factorizations. We also plan to apply it to libraries of other proof assistants assuming we can get the dependency information which axioms/theorems have been used in which proof. 
Also other automation tactics and especially
heuristics to identify isomorphic formulas need to be explored, as
well as heuristics to identify subsets for horizontal and vertical
splits. On a more theoretical level, we will investigate how axioms
and theorems could be identified, in order to allow to factorize
alternative axiomatizations of the same theory without losing
information, such as, e.g., alternative forms to axiomatize
groups. Finally, the whole system can be applied to any untyped
first-order subset of TPTP theories to search for
redundancies. However, the resulting development graphs cannot be
saved as TPTP theories, as it does not support renaming. Hence, we
propose to extend the TPTP language in that respect.

\bibliographystyle{abbrv}

\appendix

\section*{Proof of Theorem \ref{Thm:Soundness} (Structure Preservation)}
\label{sec:soundness proofs}

\subsection*{Horizontal Split}

  It holds trivially that $\Dom{\DG} = \Dom{\DG'}$. 
  \begin{itemize}
	
  \item $\loc'$ is surjective because by construction each $N_i$, $i=1,\ldots,k$ has a local entity.
    Furthermore, for each $N_i$ and each $e\in\dom^{N_i}$ holds $\loc'(e) = N_i$ by construction.
    Furthermore, since $\loc^{-1}(N) = \dom^N$, none of the incoming links into $N$ provided any 
		entity, and consequently none of the incoming links into $N_1, \ldots, N_k$ do.
    Hence, $\loc'^{-1}(N_i) = \dom^{N_i}$, $i=1,2$ and since 
		$\dom^N := \dom^{N_1}\uplus\ldots\uplus \dom^{N_k}$, $\loc'(e)$ is unique for $e\in\dom^N$.
		
  \item If $N$ is not a top-level node in $\DG$, then 
	  $\Dom{\DGRoots{\DG'}} = \Dom{\DGRoots{\DG}} = \Sigma\uplus \AllAxioms\uplus \AllLemmata$ 
		because the domains of nodes reachable from $N$ are not affected by the horizontal split.
    If $N$ is a top-level node, then all $N_i$ with $1 \le i \le k$ are top-level nodes.
    Since $dom^N = \dom^{N_1}\uplus \ldots \uplus \dom^{N_k}$ and 
		      $\Imports{\DG}(N) = \Imports{\DG'}(N_1) = \ldots = \Imports{\DG'}(N_k)$, it holds
    \begin{eqnarray*}
      \Dom{\DG}(N) & = & \dom^N\cup\Imports{\DG}(N) = 
       % \\ & = &
       \dom^{N_1}\cup \ldots\dom^{N_k} \cup \Imports{\DG}(N)\\
      & = & \dom^{N_1}\cup \ldots\dom^{N_k} \cup \Imports{\DG'}(N_1) \cup\ldots \cup \Imports{\DG'}(N_k)  \\ & = & 
      \dom^{N_1}\cup \Imports{\DG'}(N_1) \cup \ldots \cup \dom^{N_k}\cup\Imports{\DG'}(N_k)   \\ & = &
       \Dom{\DG'}(N_1) \cup \ldots \cup \Dom{\DG'}(N_k)
    \end{eqnarray*}
    Thus, $\Dom{\DGRoots{\DG'}} = \Dom{\DGRoots{\DG}} = \Sigma\uplus \AllAxioms\uplus \AllLemmata$.
		
  \item Assume $\phi\in\Lemmas{\DG}$ and $\psi\in\Support(\phi)$.
    If $\loc_{\DG}(\psi) \neq N$ and $\loc_{\DG}(\phi) \neq N$, then both 
		$\loc_{\DG}(\psi), \loc_{\DG}(\phi)$ are in $\DG'$ and we consider 
		$p:\loc_{\DG}(\psi)\greaches{\presigma}\loc_{\DG}(\phi)$.
    If $N\in p$ then $p := [p_1, \glinka{M}{\pretheta}{N}\glinka{}{\pretau}{M'}, p_2]$ 
		and by construction the path 
		$[p_1, \glinka{M}{\pretheta}{N_i}\glinka{}{\pretau_{|\Dom{N_i}}}{M'}, p_2]$ are in $\DG'$ for $1 \le i \le k$.
    Since $\loc_{\DG}(\psi) \neq N$, each $\pretau_{|\Dom{N_i}}$ behaves equivalently on the image of $\psi$ imported in $N_i$ 
		and hence $\loc_{\DG'}(\psi)\greaches{\presigma'}\loc_{\DG'}(\phi)$ for some $\presigma'$ 
		such that $\presigma'(\psi) = \presigma(\psi)$.
    If $N\not\in p$, then $p$ is also a path in $\DG'$ and 
		$\loc_{\DG'}(\psi)\greaches{\presigma}\loc_{\DG'}(\phi)$ holds trivially.
    
    If $\loc_{\DG}(\phi)=N$ then since all $N_i$ are mutually lemma independent, 
		without loss of generality we can assume $\phi\in\axioms^{N_1}\cup\lemmas^{N_1}$ and this $\loc'_{\DG'}(\phi) = N_1$.
    If $\loc_{\DG}(\psi) = N$, then $\psi'\in\axioms^{N_1}\cup\lemmas^{N_1}$ because $N_1$ is lemma independent.
    Thus, $\loc'_{\DG'}(\psi) = N_1$ and $\loc'_{\DG'}(\psi)=N_1\greaches{id}N_1=\loc'_{\DG'}(\phi)$ holds trivially.
    Otherwise, $\loc_{\DG}(\psi) = \loc'_{\DG'}(\psi)$ and since $N$ was reachable from $\loc_{\DG}(\psi)$ 
		by construction $N_1$ is also reachable from $\loc'_{\DG'}(\psi)$.
  \end{itemize}

\subsection*{Vertical Split}
\begin{itemize}
\item First, we have to prove that $\loc'$ is a location mapping.
  $\loc'$ is surjective because by construction each node $N_i$ (with $i = 1,2$) has some local entity $e \in \dom^{N_i}$.
  Thus $\loc'(e) = N_i$ and $N_i$ is in the range of $\loc'$.
  Furthermore, $\forall e\in \dom^{N_i}.\; \loc'(e) = N_i$ holds by definition.
  Finally, let $e \in \Dom{\DG'} = \Dom{\DG}$:
  $\loc'(e) = N_i$ implies $\loc(e) = N$ and therefore there is no node in $\calN \setminus \{N\}$ which provides $e$.
  Furthermore, since $\glinka{N_1}{id}{N_2} \in \calL'$, $N_1$ and $N_2$ cannot provide the same entity $e$.
\item By definition $\forall e \in \dom^{N_i}$ implies $\loc'(e) = N_i$ for $i = 1,2$ in $\DG'$.
  For all other nodes in $\DG' \setminus \{N_1, N_2\}$ the property is inherited by $(\DG, \loc, \Support)$ being 
	a structuring and $\loc(e) = \loc'(e)$ if $\loc(e) \not= N$.
\item Since $\Dom{\DG}(N) = \Dom{\DG'}({N_2})$ and $N \greaches{\presigma}{M} \in \DG$ iff $N_2 \greaches{\presigma}{M} \in \DG'$ $\Dom{\DGRoots{\DG}} = \Dom{\DGRoots{\DG'}}$.
\item Suppose $\phi \in \Lemmas{\DG}, \psi \in \Support(\phi)$ with $\loc(\phi) = M$ and $\loc(\psi) = M'$.
  If $N \not\in \{M, M'\}$ then $\loc'(\phi) = M$, $\loc'(\psi) = M'$ and $M \greaches{\presigma} M'$ in $\DG'$ trivially.
  If $M = N$ and $M' \not= N$ then $\loc'(\phi) \in \{N_1, N_2\}$, and again $N_i \greaches{\presigma} M'$ in $\DG'$.
  The case of $M \not= N$ and $M' = N$ is proven analogously.
  We are left with the case of $M = M' = N$.
  
  Since $N_1$ is independent of $N_2$ , it holds that for all $\phi' \in \axioms^{N_1} \cup \lemmas^{N_1}.
  \; \Support(\phi) \cap (\axioms^{N_2} \cup \lemmas^{N_2}) = \emptyset$.\vspace*{-1ex}\par
  \noindent Thus $\phi \in \axioms^{N_1} \cup \lemmas^{N_1}$ implies that $\psi \in \axioms^{N_1} \cup \lemmas^{N_1}$ as well and $N_1 \greaches{id} N_1$ holds trivially.
  \qed
\end{itemize}

\subsection*{Factorization}

 \begin{itemize}
  \item We have to prove that $\loc'$ is a location mapping. 
	First, we prove that $\loc'$ is surjective. For any node $K \in \calN' \setminus \{N, N_1, \ldots N_p\}$
	$\loc^{-1}(K) = \loc^{-1}(K)$ holds. Since $\signature^N \cup \axioms^N \not= \emptyset$ but 
	$(\signature^N \cup \axioms^N) \cap \Dom{\DG} = \emptyset$ it holds that $\signature^N \cup \axioms^N \subseteq \loc'^{-1}(N)$. 
	Furthermore, $\signature^{M_j} \cup \axioms^{M_j} \subseteq \loc'^{-1}(N_j)$ 
	since $\signature^{M_j} \cup \axioms^{M_j} \subseteq \pretheta_j(\signature^N \cup \axioms^N)$ and
	$\pretheta_j(\signature^N \cup \axioms^N) \cap (\signature^N \cup \axioms^N)  = \emptyset$.
	
	Second we have to prove $\forall K\in \calN'. \; \forall e \in \dom^K. \; \loc'(e) = K$ holds.
	If $K \not\in \{N, N_1, \ldots N_p\}$ then $\loc'(e) = \loc(e) = K$. If $K = N$ then $\dom^N \in \Dom{\DG'}(N)$
	and $\dom^N \not\in \Dom{\DG}(K_i)$ for $i = 1,\ldots, n$ because $\dom^N \cap \Dom{\DG} = \emptyset$. Thus 
	$\forall e \in \dom^N. \; \loc'{\DG'}(e) = N$. Finally, if $K = N_j$ then $\dom^{N_j} = \lemmas^{M_j} \setminus \theta_j(\lemmas)$
	In particular, $\dom^{N_j} \cap \Dom{\DG'}(N) = \emptyset$ implying that $\loc'{\DG'}(e) = N_j$ for all $e \in \dom^{N_j}$.
	
	Third, we prove that all $e \in \Dom{DG'}$ are provided by a unique node. The only interesting case is that $e$ is provided by
	$N$ or some $N_j$. In case of $N$ both $\dom^N$ and also entries provided by some link from $K_i$ are by definition not in $\Dom{\DG}$
	and thus not provided by any node already in $\DG$ but by definition also not provided by $N_j$. It remains the case that an entry $e$
	is provided by two nodes $N_i$ and $N_j$. Since all $e \in \Dom{DG}$ were provided by a unique node, this implies that $e$ has to be a mapped
	lemma of $N$ but that violates the precondition that each $\theta_i$ has to map $e$ into a different entity.
		
  \item Next we prove that $\DG$ and $\DG'$ coincide in the entities they provide at their maximal nodes. 
	Since $N$ is not a maximal node, it is sufficient to prove that $N_j$ and $M_j$ coincide in their provided entities: 
	\begin{equation*}
	 \begin{split}
	 \Dom{\DG'}(N_j)   
	 & =  \lemmas^{M_j} \setminus \theta_j(\lemmas) \; \cup \; 
	   \bigcup \{\sigma(\Dom{\DG'}(K)) \;|\; \glinka{K}{\presigma}{N_j} \}  \\
	 & =  \lemmas^{M_j} \setminus \theta_j(\lemmas) \; \cup \; 
	   \bigcup \{\sigma(\Dom{\DG'}(K)) \;|\; \glinka{K}{\presigma}{N_j}, K\not=N \} \\
	 &  \quad  \cup \; \theta_j(\signature) \cup \theta_j(\axioms) \cup \theta_j(\lemmas) 
	   \; \cup \; \bigcup \{ \sigma_{i,j}(\Dom{\DG}(K_{i,j})) | i=1...n \} \\
	 & =  \lemmas^{M_j} \cup \signature^{M_j} \cup \axioms^{M_j} \; \\
	 & \quad \cup \; 
	     \bigcup \{\sigma(\Dom{\DG}(K)) \;|\; \glinka{K}{\presigma}{M_j}, K\not=K_i, \sigma\not=\sigma_{i,j} \} \\
	  & \quad  \cup \; \bigcup \{ \sigma_{i,j}(\Dom{\DG}(K_{i,j})) \;|\; i=1...n\} \; 
	     \cup \; \theta_j(\lemmas) \\
     & =  \Dom{\DG}(M_j) \cup \theta_j(\lemmas).
     \end{split}
     \end{equation*}
	
	\item Suppose $\phi \in \Lemmas{\DG'}$ and $\psi \in \Support_{\DG'}(\phi)$. 
	If $\loc'(\phi), \loc'(\psi) \not\in \{N, N_1,\ldots N_p\}$ then $\loc'(\phi) = \loc(\phi)$ and
	$\loc'(\psi) = \loc(\psi)$ and therefore, $\exists \presigma . \; \loc(\psi) \greaches{\presigma} \loc(\phi)$ with
	$\presigma(\psi) = \psi$ in $\DG$. Since $\DG'$ inherits all links away from $M_1, \ldots M_p$ and paths 
	travesing some $K_i$ and $M_j$ can be mapped to paths traversing $K_i$, $N$, and $N_j$. 
	$\exists \presigma . \; \loc'(\psi) \greaches{\presigma} \loc'(\phi)$ with $\presigma(\psi) = \psi$ also in $\DG'$-
	Next, let $\loc'(\phi) = N_j$: by definition we know that $\phi \in M_j$ and 
	$\Support(\phi) \subseteq \Dom{\DG}(M_j)$. Since $\Dom{\DG}(M_j) \subseteq \Dom{\DG'}(N_j)$ we know that 
	$\Support'(\phi) = \Support(\phi) \subseteq \Dom{\DG'}(N_j)$ and thus $\forall \psi \in \Support'(\phi). \;
	\loc'(\psi) \greaches{\presigma}{N_j}$ with $\presigma(\psi) = \psi$.
	Finally, let $\loc'(\phi) = N$. Then $\Support_N \subseteq \Support'$ is a support mapping for $\phi$ in
	particular. 
 \end{itemize}

\subsection*{Transitive enrichment}
Obviously, the inclusion of the global link does not affect the visibility (e.g. $\Dom{}$) 
of any node in $\calN$ nor the local entities provided by the individual nodes (i.e. $\dom$).
Hence, all properties of a structuring are trivially forwarded to the enriched structuring.

\subsection*{Removable link}
\begin{itemize}
  \item We have to prove that $\loc$ is also a location mapping for $\DG'$. 
	It holds that $\forall N\in \calN.\; \loc_{\DG}(N) = \loc_{\DG'}(N)$ since $\dom(N)$
	remains unchanged and also all $e \in \loc_{\DG}(N)$ that are exclusively provided by some
	link in $\DG$ are still provided exclusively in $\DG'$. Thus, $\loc$ is also surjective in
	$\DG'$, also $\forall N \in \calN. \forall e\in\dom^{N}. \; \loc{\DG'}(e) = \loc{\DG}(e) = N$
	and $\forall e \in \Dom{\DG'}. \; \loc{\DG'}(e)$ is the only node providing $e$.
	\item $\DG'$ and $\DG'$ coincide in the entities they provide at their maximal nodes, which is
	 an immediate consequence of condition (2) of Def. \ref{def:remlink}.
	\item Also $\forall \phi\in\Lemmas{\DG'}\;.\; \forall \psi\in\Support(\phi).\; 
	  \exists \presigma.\;\loc(\psi) \greaches{\presigma} \loc(\phi) \wedge \presigma(\psi) = \psi$ is
		implied by condition (3) of Def. \ref{def:remlink}.
\end{itemize} \qed

\end{document}

%% file: defs.tex
\newcommand{\calN}{\mathcal{N}}

\newcommand{\ispan}[5]%
  {{#1}\stackrel{#2}{\longleftarrow}{#3}\stackrel{#4}{\longrightarrow}{#5}}
\newcommand{\isink}[5]%
  {{#1}\stackrel{#2}{\longrightarrow}{#3}\stackrel{#4}{\longleftarrow}{#5}}

%% file: results.tex
\verb|binop_2.top.rated| & 21 / 19 & 28 / 28 &5\% &   yes \\
\verb|bintree1.top.rated| & 62 / 61 & 16 / 16 &2\% &   no \\ 
\verb|cfuncdom.top.rated| & 25 / 24 & 40 / 40 &2\% &   no \\ 
\verb|ff_siec.top.rated| & 52 / 51 & 32 / 32 &2\% &   no \\ 
\verb|finsub_1.top.rated| & 38 / 37 & 16 / 16 &2\% &   no \\ 
\verb|heine.top.rated| & 96 / 95 & 13 / 13 &1\% &   no \\ 
\verb|membered.top.rated| & 17 / 17 & 36 / 16 &38\% &   no \\ 
\verb|mssubfam.top.rated| & 84 / 83 & 55 / 55 &1\% &   no \\ 
\verb|msualg_1.top.rated| & 49 / 48 & 13 / 13 &2\% &   no \\ 
\verb|power.top.rated| & 103 / 102 & 61 / 61 &1\% &   yes \\
\verb|qc_lang1.top.rated| & 86 / 85 & 23 / 23 &1\% &   no \\ 
\verb|rsspace.top.rated| & 46 / 45 & 20 / 20 &2\% &   no \\ 
\verb|setfam_1.top.rated| & 51 / 48 & 44 / 44 &4\% &   no \\ 